# Critical state model for explosive emission plasma parameters estimation

M.M. Tsventoukh

Lebedev Physical Institute of Russian Academy of Sciences

**Abstract**

A model has been proposed for estimation of plasma parameters of explosive electron emission pulses in vacuum arc discharge. It based on transition through the critical state during the explosion and allow to predict the cathode spot plasma parameters for various materials. The cathode flare plasma ions kinetic energy was estimated to be of about 100 times critical temperature. Average ions charge has been estimated as (1 + critical temperature value in eV). Both dependencies agree with experimental results. Explosive electron emission plasma momentum per transferred charge has been evaluated to be about tens of g cm / (s C) and agrees with the product of measured ions velocity and erosion rate. Effective critical temperature approach has been proposed for the estimation of plasma properties for the arc burning at surface with a fine structure.

## Introduction

Vacuum arc discharge implies plasma formation from the electrode material [1-2]. Explosive electron emission (EEE) splashes – ectons provide electron and plasma emission from the cathode [3]. Fast (ns) operation of multiple microexplosions makes it difficult to perform complete modelling of their ensembles forming the cathode spot. In addition to the detailed models [4-6] we have developed a simple approach based on the consideration of transition through the critical state of matter during the EEE pulse [7-8]. It corresponds to temperature and density of about 1 eV and $10^{22}$ cm$^{-3}$ [9-10]. Note that minimum in conductivity and hence maximum in Joule energy release also arise at about 1 eV [11].

This approach allows to evaluate general features of the EEE plasma for various materials. Its application to the liquid-metal jet tearing and electrical explosion problem allowed to evaluate the plasma and current density, plasma momentum and ohmic electric field that agrees with measurements and supports pressure-driven plasma acceleration model and ohmic nature of the cathode potential fall [8].

This approach would be helpful for estimations of erosion rates at fusion devices first wall [12-15] and for plasma composition during arc burning at complicated multicomponent cathodes [16-20].

Recently new experimental results have been obtained for vacuum arc burning at surface with a fine structure. Variation in the ions charge state, their kinetic energy and arc burning voltage has been observed. The aim of the present work to extend the critical state model to describe the observed dependencies.

Table 1 presents known vacuum arc plasma parameters for the set of materials used further.

Table 1. Parameters of vacuum arc plasma ($T_{cr}$, $v_i$, $\gamma_i$, $\alpha_i$, $Z_i$, $U_c$, $E_{coh}$) taken from [1, 2, 9, 21, 22]

|    | z  | $M_i/M_p$ | $T_{cr}$, K | $T_{cr}$, eV | $\gamma_i$, µg/C | $v_i$, 1e6 cm/s | $\alpha_i$, % | $Z_i$ | $U_c$ | $E_{coh}$ |
|----|----|-----------|-------------|--------------|------------------|-----------------|---------------|-------|-------|-----------|
| Li | 3  | 6.941     | 3223        | 0.28         |                  | 2.31            | 10            | 1     | 23.5  | 1.63      |
| Al | 13 | 26.98     | 8000        | 0.69         | 15.9             | 1.54            | 11.2          | 1.7   | 23.6  | 3.39      |
| Ti | 22 | 47.88     | 11790       | 1.02         | 22.4             | 1.54            | 9.7           | 2.1   | 21.3  | 4.85      |
| Fe | 26 | 55.8      | 9600        | 0.83         | 50               | 1.26            | 10            | 1.82  | 22.7  | 4.28      |
| Co | 27 | 58.93     | 10460       | 0.90         | 30.4             | 1.21            | 9.6           | 1.7   | 22.8  | 4.39      |
| Cu | 29 | 63.54 | 8390 | 0.72 | 33.4 | 1.32 | 11.4 | 2.06 | 23.4 | 3.49 |
| Zr | 40 | 91.22 | 16250 | 1.40 | 36.3 | 1.54 | 10.5 | 2.58 | 23.4 | 6.25 |
| Cd | 48 | 112.4 | 2790 | 0.24 | 94.6 | 0.68 | 12 | 1.32 | 16 | 1.16 |
| In | 49 | 114.8 | 6120 | 0.53 | 80.5 | 0.6 | 10.2 | 1.34 | 17.5 | 2.52 |
| Sn | 50 | 118.7 | 8200 | 0.71 | 83.1 | 0.7 | 11.4 | 1.53 | 17.5 | 3.13 |
| Sm | 62 | 150.4 | 5340 | 0.46 | 46.1 | 0.81 | 6.5 | 2.13 | 14.6 | 2.14 |
| Ta | 73 | 180.9 | 20570 | 1.77 | 31.2 | 1.2 | 5.3 | 2.93 | 28.7 | 8.1 |
| W | 74 | 183.9 | 21010 | 1.81 | 27.1 | 1.11 | 5 | 3.07 | 31.9 | 9 |
| Pt | 78 | 195.1 | 14330 | 1.24 | 50.6 | 0.81 | 5.6 | 2.08 | 22.5 | 5.84 |
| Pb | 82 | 207.2 | 4980 | 0.43 | 173 | 0.58 | 14.3 | 1.64 | 15.5 | 2.03 |
| Bi | 83 | 209 | 4200 | 0.36 | 172 | 0.47 | 10.2 | 1.17 | 15.6 | 2.18 |

**Drift velocities of ions and electrons in cathode flare plasma**

The cathode spot of the vacuum arc emits plasma jet due explosive splashing. The ions velocity have been measured in details [2]. It is worth noting that the mean ions velocity vary just in the small range of about 5 to 20 km/s for all of the conditions.

According to the critical state model, one would attribute this almost constant velocity to the critical state parameters. Indeed on can see from Table 2 that the ions velocity $v_i$ is about tenfold larger than ion acoustic velocity $(T_{cr}/M_i)^{1/2}$ :

$$v_i \sqrt{\frac{M_i}{T_{cr}}} \approx 11. \tag{1}$$

Electron drift velocity $v_e$ estimated as $v_i/\alpha_i$ [8] ($\alpha_i$ is the fraction of ion current [1-2]) corresponds to $(T_{cr}/2\pi m_e)^{1/2}$ (see Table 2):

$$\frac{\alpha_i}{v_i} \sqrt{\frac{T_{cr}}{2\pi m_e}} \approx 1. \tag{2}$$

Table 2. Ion and electron drift velocities

| | $M_i/M_p$ | $T_{cr}$, eV | $v_i$, 1e6 cm/s | $\alpha_i$, % | $v_i (T_{cr}/M_i)^{1/2}$ | $\alpha_i/v_i (T_{cr}/2\pi m_e)^{1/2}$ |
|---|---|---|---|---|---|---|
| Li | 6.941 | 0.28 | 2.31 | 10 | 11.54 | 0.39 |
| Al | 26.98 | 0.69 | 1.54 | 11.2 | 9.63 | 1.03 |
| Ti | 47.88 | 1.02 | 1.54 | 9.7 | 10.57 | 1.08 |
| Fe | 55.8 | 0.83 | 1.26 | 10 | 10.34 | 1.23 |
| Co | 58.93 | 0.90 | 1.21 | 9.6 | 9.78 | 1.28 |
| Cu | 63.54 | 0.72 | 1.32 | 11.4 | 12.37 | 1.25 |
| Zr | 91.22 | 1.40 | 1.54 | 10.5 | 12.42 | 1.37 |
| Cd | 112.4 | 0.24 | 0.68 | 12 | 14.70 | 1.47 |
| In | 114.8 | 0.53 | 0.6 | 10.2 | 8.85 | 2.10 |
| Sn | 118.7 | 0.71 | 0.7 | 11.4 | 9.07 | 2.33 |
| Sm | 150.4 | 0.46 | 0.81 | 6.5 | 14.64 | 0.93 |
| Ta | 180.9 | 1.77 | 1.2 | 5.3 | 12.12 | 1.00 |
| W | 183.9 | 1.81 | 1.11 | 5 | 11.18 | 1.03 |
| Pt | 195.1 | 1.24 | 0.81 | 5.6 | 10.18 | 1.31 |
| Pb | 207.2 | 0.43 | 0.58 | 14.3 | 12.74 | 2.75 |
| Bi | 209 | 0.36 | 0.47 | 10.2 | 11.29 | 2.22 |

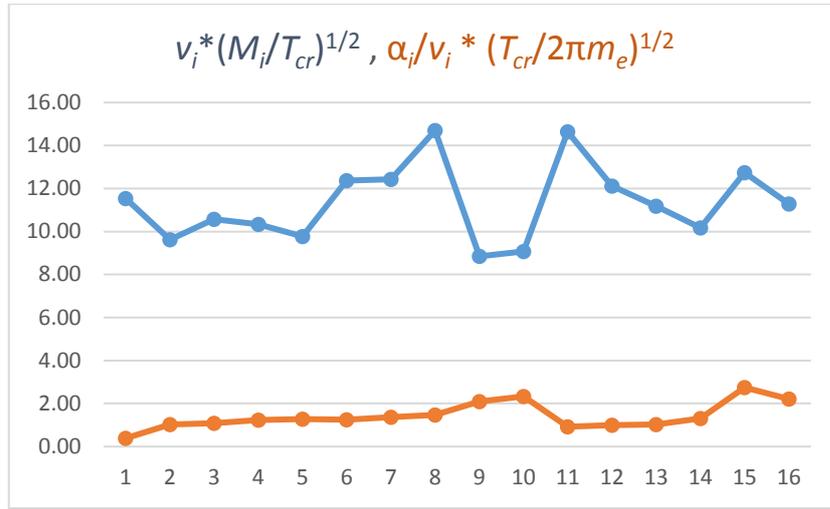

Figure 1. Ion and electron drift velocities

**Average ions charge state**

For weakly nonideal plasma average charge state $Z$ can be estimated from equation [23]

$$Z = \frac{AT^{3/2}}{n}\exp\left(-\frac{I(Z)-\delta I(n,T)}{T}\right), \qquad (3)$$

where $A = 6*10^{21}$ cm$^{-3}$ eV$^{-3/2}$, $T$ – temperature, $n$ – density. Average ionizing potential function from Z can be estimated for tungsten as $I_0 \approx I_1 * Z \approx 8$ eV $* Z$. The reduction of ionizing potential due to the Coulomb interaction can be estimated as $\delta I = Ze^2/L_D$, where $L_D = (T/4\pi Z^2 e^2 n)^{1/2}$ – Debay length. Resulting equation for ionizing potential then reads

$$I(Z) - \delta I(n,T) = Z\,I_1 - Z^2 e^2 \sqrt{\frac{4\pi e^2 n}{T}} = 8\text{ eV}*Z - 1.9\text{e-}10\text{ eV}^{\frac{1}{2}}\text{ cm}^{3/2}*Z^2\sqrt{\frac{n}{T}}. \qquad (4)$$

The plasma density decreases strongly from $10^{22}$ to $10^{18}$ cm$^{-3}$ during the explosive plasma expansion, and average plasma density has been estimated to be about $10^{20}$ cm$^{-3}$ [8].

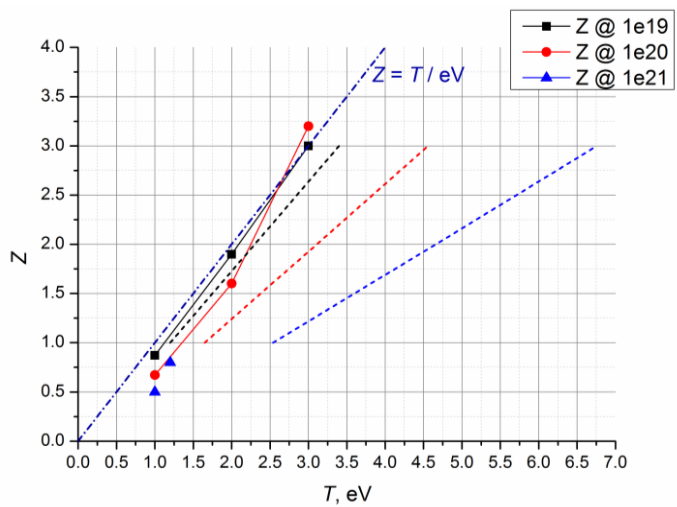

Figure 2. Average charge dependence from the temperature for various plasma density

One may find from (3) with (4) that the average charge state can be estimated as [12]

$$Z \approx T/\text{eV}. \qquad (5)$$

This agrees with more detailed models of nonideal arc spot plasma charge state composition [24] and Saha freezing model [25], see also Ref. 26.

**Electron and ion temperature calculations**

The charge state depends on the electron temperature. For fast sub-nanosecond time scales, it may differ from the ion temperature due to finite relaxation rate. In particular, Joule energy release results in the electrons heating, and the ions in metal receive energy by collisions (see recent example of calculations in Ref. 27).

Consider model of liquid-metal-jet tearing and electrical explosion [8]. It predicts the temperature evolution over the EEE pulse. However, it should be improved by consideration of electron and ion temperature separately due to a fast enough time scale. The electrons and ions temperatures evolution equations are

$$c_e \frac{\partial T_e}{\partial t} = \frac{j^2}{\sigma} - G(T_e - T_i) - T_e \mathit{divv}$$

$$c_i \frac{\partial T_i}{\partial t} = G(T_e - T_i) - T_i \mathit{divv}$$

The radius of a jet $r$ reduces linearly until the critical temperature is reached in the ion component $T_i = T_{cr}$. Since that moment ($t = t_{ex}$, $r = r_{0min}$) the radius $r$ increases and plasma expands spherically. Recall that transition over critical point implies the cohesive energy vanishing and strong ionizing potential reduction [25]. This allows to assume plasma formation from the liquid-metal jet neck at reaching $T_{cr}$. Radius variates as

$$\frac{dr}{dt} = \begin{cases} -v_l t, & t < t_{ex} \\ +v\, t, & t \geq t_{ex} \end{cases} \qquad (6)$$

where liquid metal velocity $v_l = 10^4$ cm/s and plasma velocity arises from pressure gradient acceleration

$$\frac{dv}{dt} = \frac{1}{M_i} \frac{T_i}{r}$$

The initial density of metal is $n_0 = 10^{22}$ cm$^{-3}$, and after $t = t_{ex}$ it reduces as $n = n_0 \, (r_{0min}/r)^2$.

The current density variates as

$$j = \frac{I}{\pi r^2(t)}$$

where $r(t)$ is given by (6). For spherical expansion $\mathit{divv}$ is

$$\mathit{divv} = \begin{cases} 0, & t < t_{ex} \\ 2v/r, & t \geq t_{ex} \end{cases}$$

Heat capacities are $c_e = T_e/\text{eV}$ and $c_i = 2.5$ before the transition to plasma state ($t < t_{ex}$) and $c_e = c_i = 3/2$ at plasma state ($t < t_{ex}$).

Energy exchange rate $G$ is given by $10^{11}$ eV$^{-1}$ s$^{-1}$ before the transition [28], and $G = 3\, m_e / M_i / \tau = 10^{-10}\, n$ cm$^3$ eV$^{-1}$ s$^{-1}$ (where $\tau = m_e^{1/2} T_e^{3/2}/(\pi \Lambda e^4 n)$ is the collisional time, $\Lambda$ is the Coulomb logarithm) [29].

Examples of calculations presented at Figures 3-4.

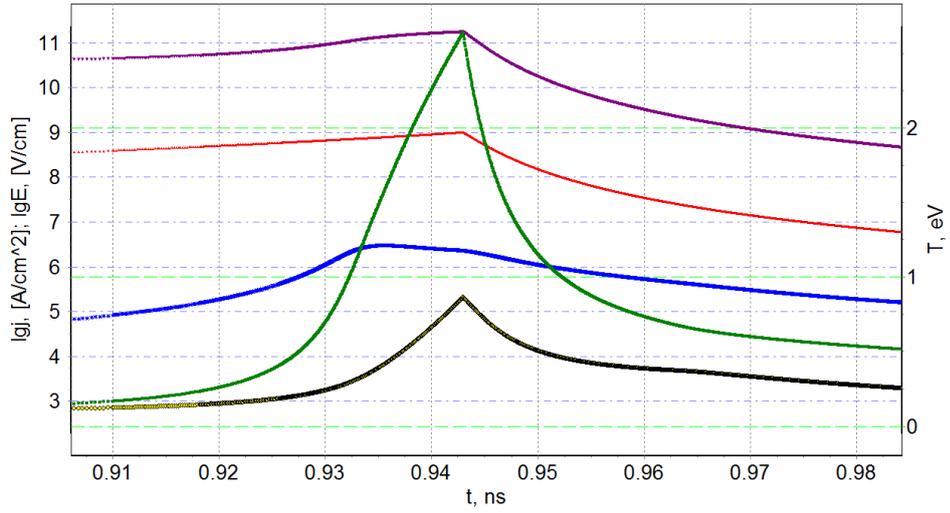

Figure 3. Fragment of temporal dependencies of $T_e$ and $T_i$, current densities $j$ and $en(T/2\pi m_e)^{1/2}$, and ohmic electric field $j/\sigma$

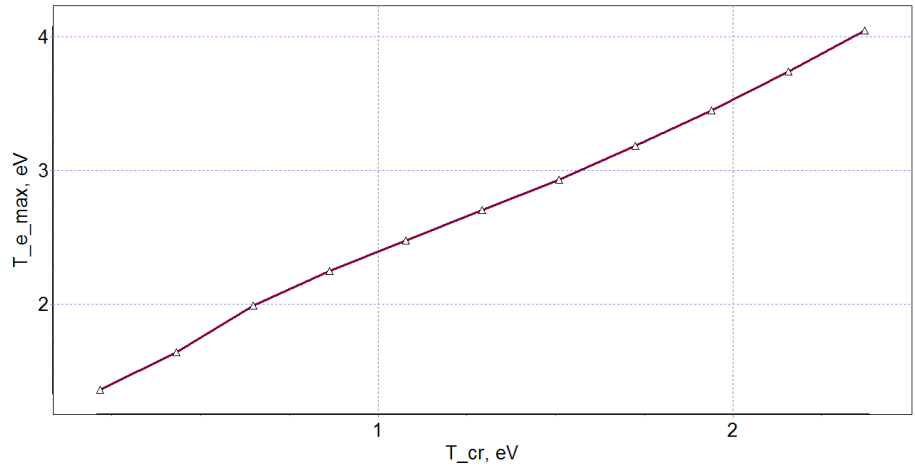

Figure 4. Dependence of maximal values of $T_e$ from critical temperature.

One may see that there is no relaxation between the electrons and ions during the fast explosion. This agrees with other results on detailed MHD and kinetic modelling [4-6].

One may find that maximal electron temperature depends from the critical one approximately as

$$T_{e,max} \approx 1 \text{ eV} + T_{cr}. \qquad (7)$$

**Average charge state estimation**

One may estimate the average charge of ions in plasma of cathode flare by Eqs (5) and (7)

$$Z_{av} = 1 + T_{cr}/\text{eV}. \qquad (8)$$

Table 3 and Figure 5 demonstrates a good agreement of the proposed formula and measurements.

Table 3. Average charge of ions

|    | $T_{cr}$, eV | $Z_i$ | $Z_{av} = 1 + T_{cr}/\text{eV}$ |
|----|---|---|---|
| Li | 0.28 | 1   | 1.28 |
| Al | 0.69 | 1.7 | 1.69 |
| Ti | 1.02 | 2.1  | 2.02 |
| Fe | 0.83 | 1.82 | 1.83 |
| Co | 0.90 | 1.7  | 1.90 |
| Cu | 0.72 | 2.06 | 1.72 |
| Zr | 1.40 | 2.58 | 2.40 |
| Cd | 0.24 | 1.32 | 1.24 |
| In | 0.53 | 1.34 | 1.53 |
| Sn | 0.71 | 1.53 | 1.71 |
| Sm | 0.46 | 2.13 | 1.46 |
| Ta | 1.77 | 2.93 | 2.77 |
| W  | 1.81 | 3.07 | 2.81 |
| Pt | 1.24 | 2.08 | 2.24 |
| Pb | 0.43 | 1.64 | 1.43 |
| Bi | 0.36 | 1.17 | 1.36 |

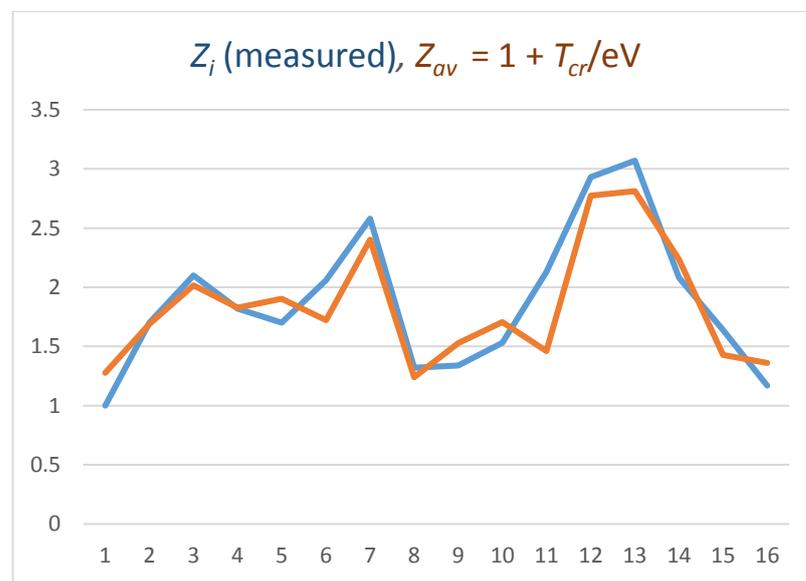

Figure 5. Average ion charge – measured and estimated by Eq. 8

**EEE-plasma momentum**

An important feature of the EEE plasma is the mechanical momentum that characterizes the acceleration rate. Momentum per transferred charge μ has been estimated to be about $(M_i T_{cr})^{1/2} / (Z_i e)$ [8]. Obtained above estimation of the average charge allows to improve this estimation as

$$\mu = \frac{\sqrt{M_i T_{cr}}}{e(1 + T_{cr}/\text{eV})}$$

As one may find from Table 4 and Figure 6 this estimation agrees with known measured values product $v_i \gamma_i$ and predicts EEE plasma momentum per transferred charge about tens of g cm/(s C).

Table 4. EEE-plasma momentum per transferred charge

|    | $M_i/M_p$ | $T_{cr}$, eV | $v_i \gamma_i$, g cm/(s C) | $\frac{\sqrt{M_i T_{cr}}}{e(1+T_{cr}/\text{eV})}$, g cm/(s C) |
|----|-----------|--------------|----------------------------|---------------------------------------------------------------|
| Li | 6.941     | 0.28         |                            | 10.87                                                         |
| Al | 26.98     | 0.69         | 24.49                      | 25.53                                                         |

| Element | | | | |
|---|---|---|---|---|
| Ti | 47.88 | 1.02 | 34.50 | 34.60 |
| Fe | 55.8  | 0.83 | 63.00 | 37.18 |
| Co | 58.93 | 0.90 | 36.78 | 38.33 |
| Cu | 63.54 | 0.72 | 44.09 | 39.34 |
| Zr | 91.22 | 1.40 | 55.90 | 47.08 |
| Cd | 112.4 | 0.24 | 64.33 | 41.92 |
| In | 114.8 | 0.53 | 48.30 | 50.95 |
| Sn | 118.7 | 0.71 | 58.17 | 53.67 |
| Sm | 150.4 | 0.46 | 37.34 | 56.98 |
| Ta | 180.9 | 1.77 | 37.44 | 64.58 |
| W  | 183.9 | 1.81 | 30.08 | 64.92 |
| Pt | 195.1 | 1.24 | 40.99 | 69.45 |
| Pb | 207.2 | 0.43 | 100.22| 65.99 |
| Bi | 209   | 0.36 | 80.61 | 63.87 |

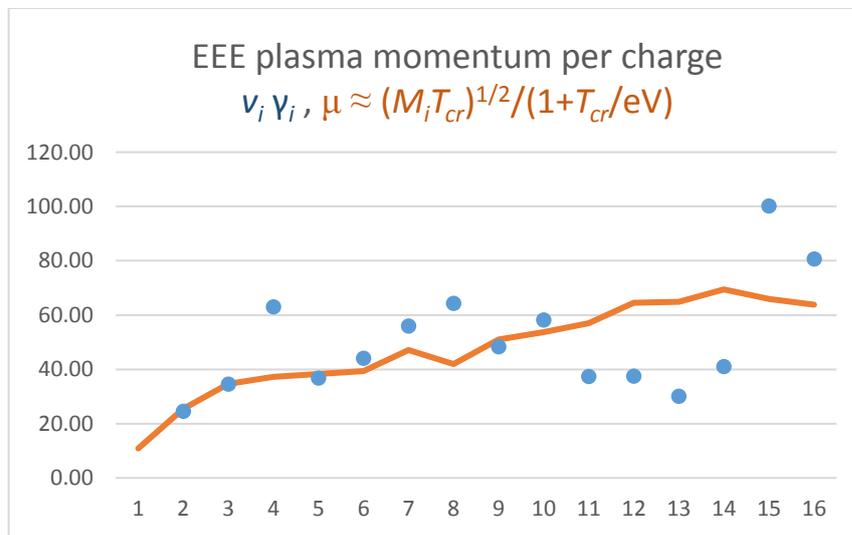

Figure 6. EEE-plasma momentum per transferred charge – experimental values $v_i \gamma_i$ and obtained equation $\mu = (M_i T_{cr})^{1/2} / e(1 + T_{cr}/eV)$

**Cohesive energy rule revisited**

It is well known that the cathode potential fall in the vacuum arc $U_c$ correlates with the cohesive energy of the cathode material $E_{coh}$ [2]. However, it depends on the energy inputted into the matter and vanishes at the critical point. So formally the cohesive energy value defined at normal state should not be spend when explosive plasma forms by transition through the critical state.

Let us use the critical temperature rather than cohesive energy. It is worth noting that the critical temperature for the materials is close to 1/5 from their cohesive energy. One may rewrite formula for the cathode fall from Ref. 30

$$U_c = U_0 + A\, E_{coh} \Rightarrow U_0 + 5A\, T_{cr}.$$

Table 5. Cathode potential fall, critical temperature and cohesive energy

|    | $T_{cr}$, eV | $U_c$ | $E_{coh}$ |
|----|------|------|------|
| Li | 0.28 | 23.5 | 1.63 |
| Al | 0.69 | 23.6 | 3.39 |
| Ti | 1.02 | 21.3 | 4.85 |
| Fe | 0.83 | 22.7 | 4.28 |
| Co | 0.90 | 22.8 | 4.39 |
| Cu | 0.72 | 23.4 | 3.49 |
| Zr | 1.40 | 23.4 | 6.25 |
| Cd | 0.24 | 16 | 1.16 |
| In | 0.53 | 17.5 | 2.52 |
| Sn | 0.71 | 17.5 | 3.13 |
| Sm | 0.46 | 14.6 | 2.14 |
| Ta | 1.77 | 28.7 | 8.1 |
| W | 1.81 | 31.9 | 9 |
| Pt | 1.24 | 22.5 | 5.84 |
| Pb | 0.43 | 15.5 | 2.03 |
| Bi | 0.36 | 15.6 | 2.18 |

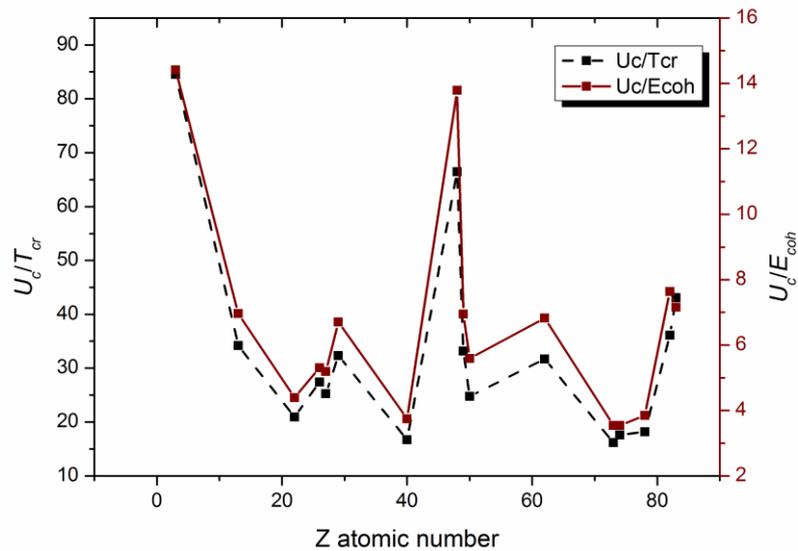

Figure 7. Ratio of cathode potential fall to the critical temperature and cohesive energy

**Effective critical temperature**

Introducing the effective critical temperature may be helpful for description of arcing at surface with a fine structure.

1. It was found recently that the average charge of ions reduces due to nanostructure of the surface where arc spot operates [12]. Arc was ignited at tungsten "fuzz" surface with nanowires layer of a micron depth. The burning at very top of the layer results in just single-charged ions observation. Whereas deepening of the spot into the layer results in appearance of $W^{+2}$ and $W^{+3}$ ions. Finally after burning out of the nanowires layer the average charge rises to that on common tungsten – about +3. Proposed reduction of the critical temperature for weakly bounded nanowire fragments allows to describe observed reduction of the ion charge by Eqs. (3) – (5).

2. Recently a deviation from cohesive energy rule has been reported in Ref. 16, where intermetallic and composite cathodes from Nb and Al have been used in vacuum arc experiments. Experiments demonstrate:

- The burning voltage was less than that should be expected from cohesive energies considerations.

- The ions kinetic energy in composition was less than that for clean materials – Nb and Al.

Taking into account Eq. (1) implying that plasma ions average kinetic energy proportional to the critical temperature

$$\frac{1}{2} M_i v_i^2 \sim 100 \, T_{cr},$$

one may propose an effective reduction of the critical temperature of the composition due to microrelief formation. Similarly the reduction of the burning voltage could be associated with reduction of the critical temperature for the microstructure.

3. In experiments of arcing at Ti – Al cathodes Ref. 19 reduction of the average charge has been obtained for the composition. In accordance with Eq. 8 it may be explained as effective critical temperature reduction due to microstructure of surface.

**Conclusions**

A model has been proposed for estimation of plasma parameters of explosive electron emission pulses in vacuum arc discharge. It based on transition through the critical state during the explosion and allow to predict the cathode spot plasma parameters for various materials from known critical temperature.

The cathode flare plasma ions kinetic energy was estimated to be of about 100 $T_{cr}$ that agrees with the result of known experiments.

The electron drift velocity in the cathode flare was estimated to be about $(T_{cr}/2\pi m_e)^{1/2}$.

Average ions charge dependence from the temperature was estimated as $Z = T$/eV.

Electron and ion temperature evolution through the explosive pulse has been calculated for improved liquid-metal jet tearing and electrical explosion model. Maximal electron temperature has been estimated to be about 1 eV + $T_{cr}$.

Average ions charge has been estimated as $Z_{av}$ = 1 + $T_{cr}$/eV that agrees well with the experimental results.

This explains observed experimentally linear dependence of the average ions charge from their kinetic energy (obtained in Ref. 18) that can be written as $Z_{av} \approx$ 1 + $E_{kin}$ /100 eV.

Explosive electron emission plasma momentum per transferred charge has been evaluated to be about $(M_i \, T_{cr})^{1/2} / e(1 + T_{cr}/eV)$ that is about tens of g cm / (s C) and agrees with the product of measured ions velocity and erosion rate $v_i \, \gamma_i$.

Effective critical temperature approach has been proposed for the estimation of plasma properties for the arc burning at surface with fine structure.

**References**

1. G. A. Mesyats 2000 *Cathode Phenomena in a Vacuum Discharge: The Breakdown, the Spark, and the Arc* (Nauka, Moscow, Russia).
2. A. Anders, 2008 *Cathodic Arcs. From Fractal Spots to Energetic Condensation* ( Springer Science + Business Media, LLC, New York, NY)
3. G. A. Mesyats 2013 *IEEE Trans. Plasma Sci.* **41**, 676–694

4. S. A. Barengolts, G. A. Mesyats, and D. L. Shmelev 2001 *JETP* **93**, 1065 [2001 Zh. Eksp. Teor. Fiz. **120**, 1227].
5. D. L. Shmelev and S. A. Barengolts, 2013 *IEEE Trans. Plasma Sci.* **41**(8), 1959
6. D. L. Shmelev, S. A. Barengolts, and M. M. Tsventoukh 2017 *IEEE Trans. Plasma Sci.* **45**(11), 3046
7. G. A. Mesyats and M. M. Tsventoukh, 2015 *IEEE Trans. Plasma Sci.* **43**, 3320
8. M.M. Tsventoukh 2018 *Phys Plasmas* **25** 053504
9. V. E. Fortov, I. T. Yakubov, and A. G. Khrapak, *Physics of Strongly Coupled Plasma* ( Oxford University Press, Oxford, UK, 2006).
10. V. I. Oreshkin, S. A. Barengolts, and S. A. Chaikovsky 2007 Tech. Phys. **52**(5), 642
11. A. W. DeSilva and G. B. Vunni, 2011 *Phys. Rev. E* **83**, 037402
12. S.A. Barengolts *et al* 2020 *Nucl. Fusion* **60** 044001
13. P.V. Savrukhin and E.A. Shestakov 2019 *Phys. Plasmas* **26** 092505
14. D. Hwangbo, *et al* 2018 *Contrib. Plasma Phys.* **58**(6-8), 608-615
15. S. Kajita et al 2013 *Nucl. Fusion* **53** 053013
16. Siegfried Zöhrer *et al* 2020 *Plasma Sources Sci. Technol.* **29** 025022
17. Konstantin Savkin *et al* 2019 *Plasma Sources Sci. Technol.* **28** 065008
18. Igor Zhirkov, Efim Oks, and Johanna Rosen 2015 *J. Appl. Phys.* **117** 093301
19. I. Zhirkov *et al* 2014 *J. Appl. Phys.* **115** 123301
20. I. Zhirkov, A. O. Eriksson, and J. Rosen 2013 *J. Appl. Phys.* **114** 213302
21. G. Yu. Yushkov, A. Anders, E. M. Oks, and I. G. Brown 2000 *J. Appl. Phys.* **88**, 5618
22. A. Anders, E. M. Oks, G. Y. Yushkov, K. P. Savkin, I. G. Brown, and A. G. Nikolaev 2005 *IEEE Trans. Plasma Sci.* **33**(5), 1532
23. Y.B. Zeldovich and Y.P. Rayzer 1967 Physics of Shock Waves and High Temperature Hydrodynamic Phenomena (New York: Academic)
24. A. Anders, S. Anders, A. Forster, and I. G. Brown 1992 *Plasma Sources Sci. Technol.* **1** 263
25. Anders A. 1997 *Phys. Rev. E* **55** 969–81
26. André Anders 2012 *Plasma Sources Sci. Technol.* **21** 035014
27. S. A. Barengolts, I. V., Uimanov and D. L. Shmelev 2019 *IEEE Trans. Plasma Sci* **47**(8) 3400-3405
28. H.E. Elsayed-Ali, T.B. Norris, M.A. Pessot and G.A. Mourou, 1987 *Phys Rev Lett* **58**(12) 1212
29. S. I. Braginskii, " Transport processes in a plasma," in *Reviews in Plasma Physics*, edited by M. A. Leontovich ( Gosatomizdat, Moscow, 1963; Consultants Bureau, New York, 1965), Vol. 1, pp. 205–311.
30. Anders A and Yushkov G Y 2002 *J. Appl. Phys.* **91** 4824–4832